\documentclass{ptapap}

\author{Karolina Bąkowska}[CAMK,OSU]
\author{Radosław Smolec}[CAMK]
\affil[CAMK]{Nicolaus Copernicus Astronomical Center\\
  Bartycka 18, 00--716 Warszawa, Poland}
\affil[OSU]{Fulbright Visiting Scholar, The Ohio State University\\  Dept. of Astronomy, 140 W. 18th Ave, Columbus, OH 43210, USA}

\title{RR Lyrae stars with Blazhko modulations in the OGLE collection}

\begin{document}

\maketitle

\begin{abstract}
We briefly discuss three cases of RR~Lyrae variable stars with Blazhko modulation from the OGLE Galactic bulge collection. In one of the discussed stars modulation with two periods is detected, in other star, we detect three Blazhko modulations.
\end{abstract}

Although RR~Lyrae stars are relatively simple pulsators, their behaviour is not fully understood. The biggest challenge is the Blazhko effect, long-term quasi-periodic modulation of pulsation amplitude and/or phase, observed in $\sim$40-50\thinspace per cent of RR~Lyr stars pulsating in the fundamental mode (RRab stars), e.g. \cite{Geza}. Mechanism behind the Blazhko modulation remains a mystery. For recent reviews see Smolec (these proceedings) and \cite{2014IAUS..301..241S}.

The Optical Gravitational Lensing Experiment (OGLE, \citealt{2015AcA....65....1U}) identified several thousands of RRab pulsators in the Galactic bulge \citep{2014AcA....64..177S}. We have started their analysis in search for the Blazhko modulation. For our analysis, we first selected 1167 RRab stars from the OGLE-IV Galactic bulge collection with the largest number of data points (at least 8000 measurements collected over 4 observing seasons). Because of large number of data points, the expected noise level in the Fourier transform is low, which allows us to detect the modulations of very low amplitude. 

Data are analysed using standard consecutive prewhitening technique. Significant periodicities are identified with the help of the discrete Fourier transform. Then, the corresponding frequencies and amplitudes are determined through the non-linear least-square fit of sine series to the data. In Blazhko stars, except fundamental mode frequency, $f_0$, and its harmonics, $kf_0$, close side peaks are also detected in the frequency spectrum. Together with peaks at $kf_0$, they form equidistant multiplets (triplets, quintuplets, etc.) -- Fourier representation of the modulation. Frequency separation between multiplet components corresponds to the modulation frequency, $f_{\rm B}$, and its inverse corresponds to the modulation period, $P_{\rm B}=1/f_{\rm B}$.

\begin{figure}
\centering
\includegraphics[width=.9\textwidth]{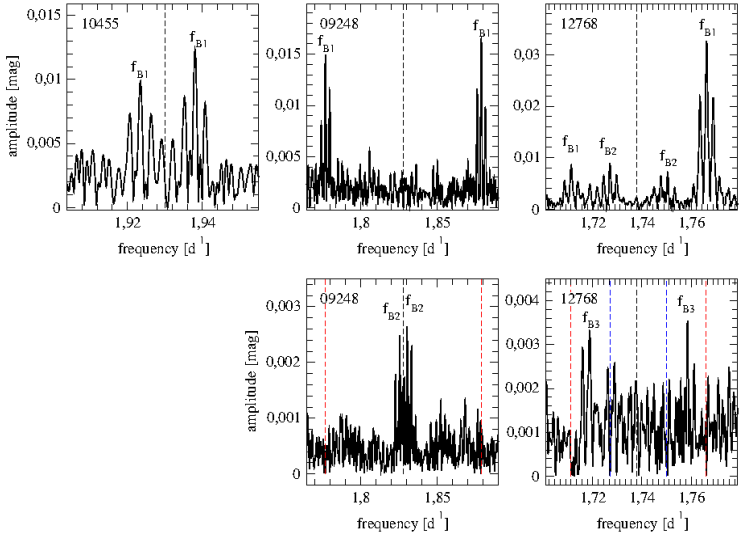}
\vspace*{-.5cm}
\caption{The prewhitened frequency spectra for 10455 (left panel), 09248 (middle panels) and 12768 (right panels). The dashed black line marks the fundamental mode frequency. In the bottom panels, red and blue dashed lines mark the location of the prewhitened Blazhko modulations, $f_{\rm B1}$ for 09248 and $f_{\rm B1}$ and $f_{\rm B2}$ for 12768.}
\label{fig:FreqSpectra}
\end{figure}

\begin{figure}
\centering
\includegraphics[width=0.85\textwidth]{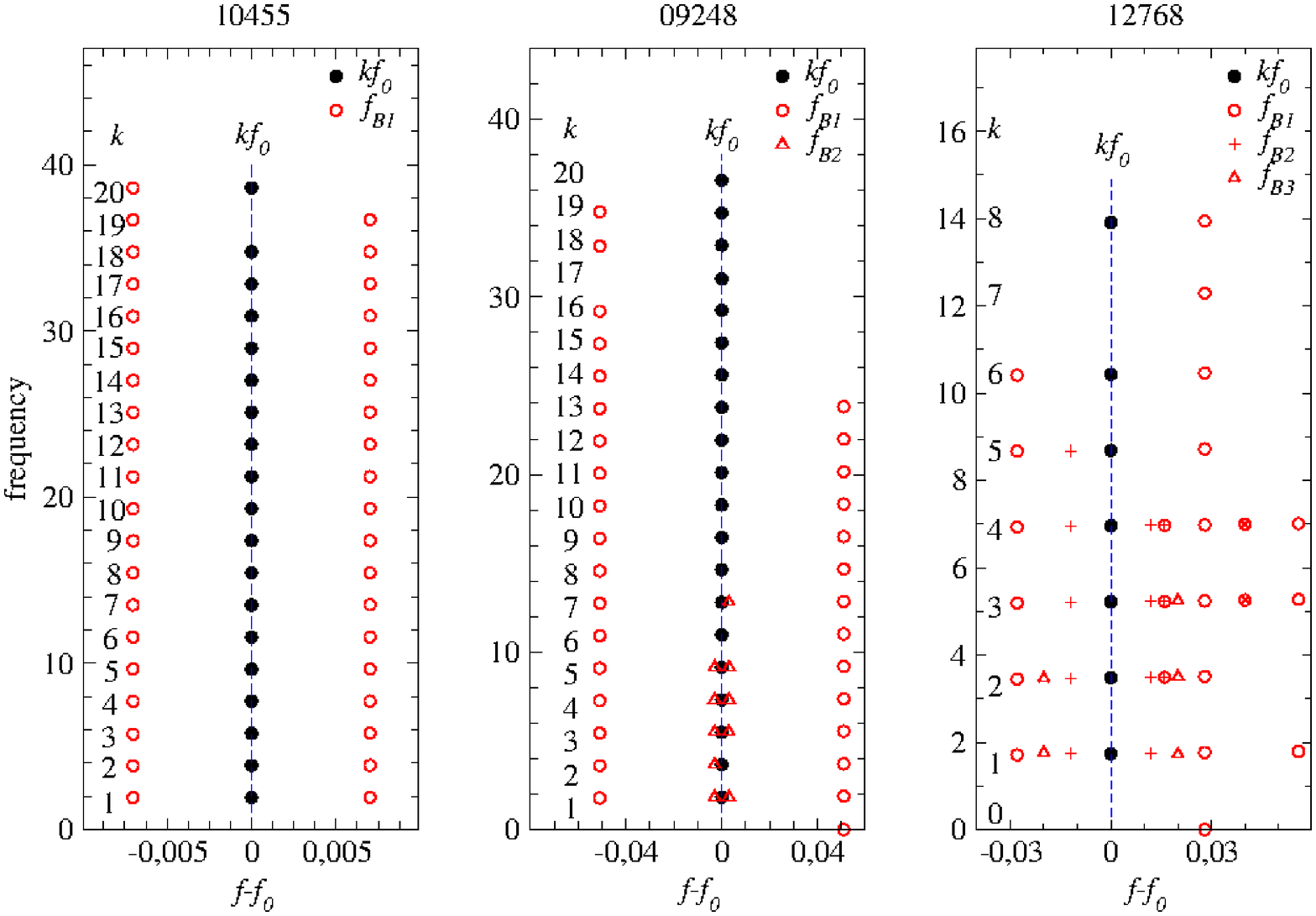}
\vspace*{-.5cm}
\caption{The {\it Echelle} diagrams for 10455 (left panel), 09248 (middle panel) and 12768 (right panel). Side peaks with different separations are marked with different symbols. Two symbols at the same location indicate the combination frequency.}
\label{fig:Echelle}
\end{figure}

Below, we present the preliminary analysis for three RRab variables with Blazhko modulations from OGLE-IV Galactic bulge collection: OGLE-BLG-RRLYR-10455, OGLE-BLG-RRLYR-09248 and OGLE-BLG-RRLYR-12768 (below denoted as 10455, 09248 and 12768, respectively). In Fig.~\ref{fig:FreqSpectra}, we show the prewhitened frequency spectra for these stars, centered on the fundamental mode frequency, $f_0$. In Fig.~\ref{fig:Echelle}, we show all peaks detected in the frequency spectra in the form of {\it echelle} diagrams, in which frequency of the detected peaks, $f$, is plotted versus $f$ modulo $f_0$ (e.g. \citealp{2012MNRAS.424..649G}). In the echelle diagrams multiplet structures are clearly visualised.

Our first object, 10455, turned out to be the most typical RRab star ($P_0\approx0.51795$\thinspace d) with single Blazhko modulation. In the prewhitened frequency spectrum in Fig.~\ref{fig:FreqSpectra} (left panel), the location of the fundamental mode frequency, $f_0$, is marked with black, dashed line and side peaks corresponding to the Blazhko modulation are labelled with $f_{\rm B1}$. The modulation period is $P_{\rm B1}=138.41$\thinspace d. In the frequency spectrum equidistant triplets are detected (Fig.~\ref{fig:Echelle}, left panel).

In the second star, $09248$ ($P_0\approx0.54702$\thinspace d), two Blazhko modulations are detected. First, modulation with shorter period,  $P_{\rm B1}=19.67$\thinspace d, was detected. In Fig.~\ref{fig:FreqSpectra} (top middle panel), the corresponding side peaks are labelled with $f_{\rm B1}$. After prewhitening, second modulation with longer period was detected (bottom middle panel in Fig.~\ref{fig:FreqSpectra}, $P_{\rm B2}\approx398$\thinspace d). For both modulations, triplet structures are present in the frequency spectrum (Fig.~\ref{fig:Echelle}).

The last case, 12768 ($P_0\approx0.57513$\thinspace d), is the most interesting one. Frequencies of the first and the second Blazhko modulations are denoted with $f_{\rm B1}$ and $f_{\rm B2}$, respectively (Fig.~\ref{fig:FreqSpectra}, top right panel). The corresponding modulation periods are $P_{\rm B1}=36.25$\thinspace d and $P_{\rm B2}=84.87$\thinspace d. After prewhitening procedure, we detect side peaks most likely corresponding to the third modulation, marked with $f_{\rm B3}$ in Fig.~\ref{fig:FreqSpectra} (bottom right panel). The corresponding modulation period is $P_{\rm B3}=50.32$\thinspace d. We note, that within frequency resolution of the data, $f_{\rm B3}=(f_{\rm B1}+f_{\rm B2})/2$. In the frequency spectrum, we detect triplet structures for all three modulations -- Fig.~\ref{fig:Echelle} (right panel). We also detect high frequency components of the quintuplet for modulation with $f_{\rm B1}$. Side peaks corresponding to the combination of the modulation frequencies are also present. Additional, low-amplitude periodicities, that cannot be interpreted as due to modulation or due to other radial modes, are also detected in 12768. 

Detailed analysis of the discussed stars, as well as full results of our survey, will be presented elsewhere. Until now, we have detected tens of RRab stars with one Blazhko modulation, several cases of RRab stars with two Blazhko modulations and, so far, only one example with three Blazhko modulations. In many cases we detect variability on a longer time scale, which cannot be fully resolved yet, due to the limited time span of observations. This variability may also correspond to Blazhko modulation of longer period. 

\acknowledgements{This project is supported by the Polish National Science Center grants awarded by the decisions DEC-2012/07/N/ST9/04172 and DEC-2015/16/T/ST9/00174 for KB and DEC-2012/05/B/ST9/03932 for RS.}

\bibliographystyle{ptapap}
\bibliography{BakowskaSmolec}

\begin{thebibliography}{5}
\providecommand{\natexlab}[1]{#1}
\providecommand{\url}[1]{\texttt{#1}}
\providecommand{\urlprefix}{URL }
\providecommand{\eprint}[2][]{\url{#2}}

\bibitem[{{Guggenberger} et~al.(2012)}]{2012MNRAS.424..649G}
{Guggenberger}, E., et~al., \emph{{The complex case of V445 Lyr observed with
  Kepler: two Blazhko modulations, a non-radial mode, possible triple mode RR
  Lyrae pulsation, and more}}, \emph{\mnras} \textbf{424}, 649 (2012),
  \eprint{1205.1344}

\bibitem[{{Kovacs}(2015)}]{Geza}
{Kovacs}, G., \emph{{The Blazhko phenomenon}}, \emph{ArXiv e-prints}  (2015),
  \eprint{1512.05722}

\bibitem[{{Soszy{\'n}ski} et~al.(2014)}]{2014AcA....64..177S}
{Soszy{\'n}ski}, I., et~al., \emph{{Over 38000 RR Lyrae Stars in the OGLE
  Galactic Bulge Fields}}, \emph{\actaa} \textbf{64}, 177 (2014),
  \eprint{1410.1542}

\bibitem[{{Szab{\'o}}(2014)}]{2014IAUS..301..241S}
{Szab{\'o}}, R., \emph{{Blazhko effect in Cepheids and RR Lyrae stars}}, in
  J.~A. {Guzik}, W.~J. {Chaplin}, G.~{Handler}, A.~{Pigulski} (eds.) IAU
  Symposium, \emph{IAU Symposium}, volume 301, 241--248 (2014),
  \eprint{1309.3969}

\bibitem[{{Udalski} et~al.(2015){Udalski}, {Szyma{\'n}ski}, \&
  {Szyma{\'n}ski}}]{2015AcA....65....1U}
{Udalski}, A., {Szyma{\'n}ski}, M.~K., {Szyma{\'n}ski}, G., \emph{{OGLE-IV:
  Fourth Phase of the Optical Gravitational Lensing Experiment}}, \emph{\actaa}
  \textbf{65}, 1 (2015), \eprint{1504.05966}

\end{thebibliography}

\end{document}